\documentclass[10pt]{article}

\usepackage[utf8]{inputenc}
\usepackage{moreverb}
\usepackage{amsmath,amsfonts,amssymb,amsthm}
\usepackage{natbib}
\usepackage{moreverb}
\usepackage{fullpage}
\usepackage{graphicx}
\usepackage{hyperref}

\newcommand\BibTeX{{\rmfamily B\kern-.05em \textsc{i\kern-.025em b}\kern-.08em
T\kern-.1667em\lower.7ex\hbox{E}\kern-.125emX}}

\newcommand{\vc}[1]{\mathbf}
\newcommand{\mat}[1]{\mathbf}

\DeclareMathOperator{\bias}{Bias}
\DeclareMathOperator{\var}{Var}
\DeclareMathOperator{\E}{\mathbb E}
\DeclareMathOperator{\diag}{diag}
\DeclareMathOperator{\mse}{MSE}
\DeclareMathOperator{\tr}{tr}
\DeclareMathOperator{\risk}{risk}
\DeclareMathOperator{\cv}{CV}

\newtheorem{definition}{Definition}

\title{Fridge: focused fine-tuning of ridge regression for personalized predictions}
\author{Kristoffer H. Hellton \&  Nils Lid Hjort \\ \\ Dept. of Mathematics, University of Oslo \\ University of Oslo, P.O.Box 1053 Blindern, 0316 Oslo, Norway. \\ E-mail: kristohh@math.uio.no }
\date{December 2017}
\begin{document}

\maketitle

\begin{abstract}
Statistical prediction methods typically require some form of fine-tuning of tuning parameter(s), with $K$-fold cross-validation as the canonical procedure. For ridge regression there exist numerous procedures, but common for all, including cross-validation, is that one single parameter is chosen for all future predictions. We propose instead to calculate a unique tuning parameter for each individual for which we wish to predict an outcome. This generates an individualized prediction by focusing on the vector of covariates of a specific individual. The focused ridge -- fridge -- procedure is introduced with a two-part contribution: 1) first we define an oracle tuning parameter minimizing the mean squared prediction error of a specific covariate vector, 2) then we propose to estimate this tuning parameter by using plug-in estimates of the regression coefficients and error variance parameter. The procedure is extended to logistic ridge regression by utilizing parametric bootstrap. For high-dimensional data, we propose to use ridge regression with cross-validation as the plug-in estimate, and simulations show that fridge gives smaller average prediction error than ridge with cross-validation for both simulated and real data. We illustrate the new concept for both linear and logistic regression models in two applications of personalized medicine: predicting individual risk and treatment response based on gene expression data. The method is implemented in the R package \verb|fridge|. \vspace{\baselineskip}
\end{abstract}

\section{Introduction}
The development of inexpensive genomic technologies has greatly contributed to the field of personalized medicine, by facilitating predictions of individualized treatment decisions and disease risks based on genetic characteristics \citep{hamburg2010path}. In Norway, for instance, the Norwegian Cancer Genomics Consortium  \href{http://cancergenomics.no}{(cancergenomics.no)} has been founded to establish ``nationwide use of individual patient genetics to guide cancer treatment''. Genomic data are typically high-dimensional with the number of variables, $p$, greatly exceeding the number of observations, $n$, and this high-dimensionality is often handled by regularization, or by constructing new, low-dimensional features \citep{hastie2001elements}. Penalized linear regression, the most widely used regularization technique, introduces some form of penalization of the regression coefficients in the linear model, $ y_{i} = x_{i}^{T}\beta +\varepsilon$. Ridge regression imposes an $L_{2}$ penalty, penalizing the sum of squared regression coefficients:
$$\hat{\beta}(\lambda) = \arg \min_{\beta} \left\{ \|Y - X \beta\|_{2}^2 + \lambda\sum_{j=1}^p \beta_{j}^2 \right\},$$
enforcing both proportional shrinkage of all coefficients towards zero (increasing the bias, but lowering the variance) and shrinking positively correlated variables towards each other. In addition, the least informative directions in the data space are penalized more, effectively improving the predictive performance of the method \citep[p.82]{hastie2001elements}. Ridge regression was originally introduced as an extension of ordinary least squares (OLS) regression to handle rank-deficient data matrices \citep{hoerl1970ridge,van2015lecture}. Ridge regression has become a standard prediction tool within genomics, and has, for instance, been shown to give the lowest error of a range of methods when predicting survival based on gene expression data \citep{bovelstad2007predicting}.

As with all penalized regression methods, the $ L_{2}$ penalty is controlled by a tuning parameter, $\lambda$, dictating the fit of the predictive machinery balancing between over- and under-fitting. A range of fine-tuning procedures has been proposed for ridge regression, including minimizing the mean squared error (MSE) of $\hat{\beta}(\lambda)$ \citep{lawless1981mean,hemmerle1975explicit}, marginal maximum likelihood \citep{tran2009penalized,johnsen2011regularized}, bootstrapping \citep{delaney1986use}, Bayesian methods \citep{zuliana2016weight} and versions of AIC \citep{boonstra2015small}. $K$-fold cross-validation (CV) has nevertheless become the canonical choice of procedure \citep[p. 243]{hastie2001elements}. CV works by dividing the data into $K$ parts, or folds, (typically 5 or 10) predicting each fold out-of-sample based on the remaining data, and finally averaging the squared prediction error over all folds. This is done for a range of tuning parameters, selecting the value with the lowest average error. Variations of CV include generalized cross-validation \citep{golub1979generalized} and approximate cross-validation \citep{meijer2013efficient}. 

Common for all procedures is that only \emph{one} tuning parameter value is found for all further use and future predictions. For some applications, however, it can be important to minimize the prediction error of a specific individual, rather than the average prediction error. In applications of personalized medicine, individual predictions can determine decisions with severe consequences: a predicted increase in the risk of complications could trigger a surveillance response, or a treatment with high predicted success probability can be initiated despite possible adverse side effects. Hence, our goal is to adjust the prediction model towards each specific patient by modifying the tuning parameter. When the risk or treatment response of a new patient is to be predicted, can we find a tuning parameter optimal for that particular patient's covariate vector $x_{0}$? Such targeting of a specific covariate vector is made possible when the optimal tuning parameter is based on minimizing the \emph{expected error} of the prediction, the so-called MSE approach. 

The selection of tuning parameter(s) shares parallels with the task of model selection, for which the focused information criterion (FIC) has introduced the concept of addressing a ‘final outcome’ of a fitted model, such as a specific prediction, instead of an overall goodness-of-fit \citep{claeskens2008model}. Along the same lines, we first define an oracle personalized tuning parameter $\lambda_{x_{0}}$, the minimizer of the expected MSE of the ridge prediction $x_{0}^T\hat\beta(\lambda)$ as a function of $\lambda$. Second, we propose to estimate the  
tuning parameter by using plug-in estimates of the regression coefficients and error variance parameters in the MSE expressions. The approach aims to minimize the expected prediction error for each individual, rather than minimizing the sample prediction error over all individuals simultaneously; in other words we average over the (theoretical) distribution of $y_{i}$ and not over the observed set of $y_{i}$s, as done by CV. Such individualized tuning parameters require a recalculation of the ridge model for each prediction, which previously would have been a constraint. But due to current computational power, this can be viewed as one approach to tailoring predictions towards the individual.  

Related work, also modifying the ridge regression tuning parameter, introduced a variable specific weight to the tuning parameter with an additional grouping determined by external data \citep{wiel2016better}. This leads to a different $\lambda$ for each (group of) covariate(s), say $\lambda_j$ for the $j$th covariate. Our approach, however, aims at producing an \emph{individualized} tuning parameter, say $\lambda(x_{0})$ for each new individual, i.e.~each vector of covariates $x_{0}$. In other words, in terms of the $n\times p$ matrix $X$, the previous work has tailored the $\lambda$ per (group of) column(s), whereas we construct one $\lambda$ per row. 

The outline of the paper is as follows: Section 2 introduces the framework of the procedure, the general definition of the oracle tuning parameter and our proposed approach to estimate the tuning parameter by plug-in estimates, and Section 3 gives an overview of the theoretical aspects of the procedure. Section 4 extends the fridge approach to the logistic regression model. Section 5 shows the results of simulations comparing the focused tuning and cross-validation in real and simulated data, and Section 6 illustrates the procedure in the linear and logistic regression models with two genomic data examples.  

\section{Method}
Consider the data $\{y_{i},x_{i}\}$, consisting of $n$ observations of a continuous outcome $y_{i}$ and $p$-dimensional vector of covariates $x_{i}$, following the linear regression model
$$y_{i}= x_{i}^T\beta+ \varepsilon_{i}, \quad i=1,\dots,n$$  
with unknown $p$-dimensional regression coefficients, $\beta$, and \emph{iid} noise, $\varepsilon_{i}$, with zero mean and variance $\sigma^{2}$. We denote the outcome vector $Y$, and the $n \times p$ data matrix $X$. For $p < n$ and a data matrix of full rank, the OLS estimate of $\beta$ is given by
$ \tilde{\beta} = (X^TX)^{-1}X^{T}Y$. 

In the high-dimensional situation with $p > n$, the least squares criterion requires a penalty to give a unique solution, and ridge regression introduces an $L_{2}$ penalty \citep{hoerl1970ridge}, also known as Tikhonov regularization, 
\begin{equation}
\hat{\beta}(\lambda) = \arg \min_{\beta} \left\{ \|Y - X \beta\|_{2}^2 + \lambda\sum_{i=1}^p \beta_{i}^2 \right\},\label{ridge} 
\end{equation}
with the \emph{explicit} solution 
$$ \hat{\beta}(\lambda) = (X^TX + \lambda I_{p})^{-1}X^{T}Y = (X^TX + \lambda I_{p})^{-1}X^{T}X \tilde{\beta}.$$
Suppose we aim to predict the expected outcome of $y_{0}$ for the vector of covariates of a specific individual $x_{0}$, the focus parameter $\mu_{0}= E y_{0}= x_{0}^T\beta$, such that the estimated ridge prediction is given 
$$ \hat\mu_{0} = x_{0}^T\hat{\beta}(\lambda) =  x_{0}^T(X^TX + \lambda I_{p})^{-1}X^{T}Y . $$
If we then consider the expected MSE of the prediction, where the expectation is taken with respect to the distribution of $Y$, the MSE will be a function of the tuning parameter $\lambda$, together with $x_{0}$ and the parameters $\beta$ and $\sigma^2$: 
\begin{align}
\mse_{\hat\mu}(\lambda;x_{0}, \beta, \sigma^2) &=  \E_{Y}\left((x_{0}^T\hat{\beta}(\lambda)-x_{0}^T\beta)^2\right)= \bias^{2}(\hat \mu) + \var \hat\mu,\nonumber\\
   &= \left\{x_{0}^{T}( (X^TX+ \lambda I_{p})^{-1}X^{T}X - I_{p})\beta \right\}^{2} \nonumber\\
   &\qquad + \sigma^{2} x_{0}^{T}(X^TX+ \lambda I_{p})^{-1}X^{T}X (X^TX+ \lambda I_{p})^{-1}x_{0}.\label{eq:mse}
 \end{align}
Note that we consider the error of $x_{0}^T\beta$ and not $y_{0}$, which simplify notation by omitting the intrinsic prediction error $\sigma^2$. 
 
\begin{figure}
\centering
 \includegraphics[width=8cm]{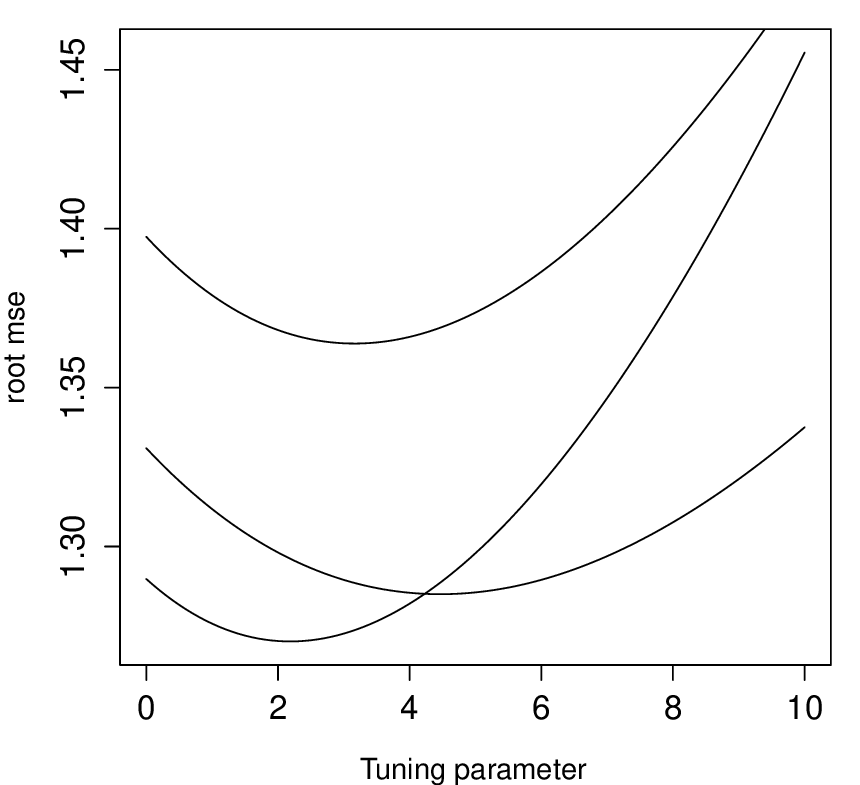}
 \caption{MSE curves for different $x_{0}$ demonstrating how the minima occur at different values of $\lambda$.}\label{fig:minima}
\end{figure}

For each specific vector of covariates $x_{0}$, i.e.~representing a new individual or patient, the MSE will have a different minimum as a function of $\lambda$, as seen in Figure \ref{fig:minima}. 
We will aim to estimate these MSE curves separately for each $x_{0}$ to lower the expected prediction error of each individual, and as this optimal value is given for known parameters $\beta$ and $\sigma^2$, it is termed the oracle tuning parameter. 
 
\begin{definition}[Oracle tuning]
The oracle tuning parameter is the minimand of the mean squared prediction error
$$ \lambda_{x_{0}} = \arg \min_{\lambda}\mse_{\hat\mu}(\lambda;x_{0}, \beta, \sigma^2), \quad \lambda \geq 0, $$
where the parameters $\beta$ and $\sigma^2$ are known. 
\end{definition}
The oracle value of the personalized tuning parameter, $\lambda_{x_{0}}$, will give the smallest expected prediction error, but cannot be used in practice as it requires the true value of $\beta$ and $\sigma^{2}$. A direct way to estimate $\lambda_{x_{0}}$ from data is to first estimate $\beta$ and $\sigma^2$ by some other method and plug-in the resulting estimates in Eq. \eqref{eq:mse}. 

In low dimension ($p < n$) the simplest choice of plug-in estimate is the ordinary least squares (OLS) estimator (assuming $X$ is of full rank), and corresponding variance estimator
$$\tilde\beta  =  (X^TX)^{-1}X^{T}Y, \quad \tilde\sigma^2 =\frac{1}{n-p}\sum_{i=1}^n(y_{i}-x_{i}^T\tilde\beta)^2.$$ 
In the high-dimensional situation ($p \gg n$), one choice of plug-in estimate is ridge regression tuned by standard CV. Alternative plug-in estimators include lasso ($L_{1}$ penalty) and principal component regression both fine-tuned by CV, or the OLS estimate combined with the Moore-Penrose pseudo-inverse. However, when ridge regression is already considered an appropriate prediction method for the problem at hand, it is natural to also use ridge as a plug-in estimate. 

The MSE in Eq. \eqref{eq:mse} relies on the squared bias, $\bias^2$, estimated by squaring the estimated bias, $(\widehat\bias)^2$, directly. The resulting estimate will be biased as $\E((\widehat\bias)^2) = \bias^2+ \var\widehat\bias,$ and a correction of the overestimation is necessary \citep[p. 150]{claeskens2008model}. We can correct the squared bias, for instance by subtracting the variance of the bias and truncate at zero
$$ \max \{ (\widehat\bias)^2- \var\widehat\bias,0\},$$
or using a smooth correction 
$$ (\widehat\bias)^2 - \frac{(\widehat\bias)^2}{(\widehat\bias)^2+1} \var\widehat\bias. $$
We will in our proposed procedure consider the first option. 

In the low-dimensional case, the OLS estimates can be used as plug-in estimates:  
\begin{definition}[Fridge-OLS]\label{fridgeOLS}
The fridge-OLS tuning parameter estimate is the minimand of the estimated mean squared error curve
\begin{align}
  \hat\lambda_{x_{0},OLS} = \arg \min_{\lambda}& \;\widehat\mse_{\hat\mu}(\lambda; x_{0},\tilde\beta,\tilde\sigma^2),\nonumber\\
   = \arg \min_{\lambda}&\left\{ \left((\widehat \bias( \lambda ))^2- \var \widehat \bias( \lambda ) \right)_{+}  + \widehat\var( \lambda ) \right\},\nonumber\\
   = \arg \min_{\lambda}& \bigg\{ \Big( (\lambda x_{0}^{T} (X^TX+ \lambda I_{p})^{-1}\tilde\beta)^{2}\nonumber\\
   &\qquad- \tilde\sigma^{2} \lambda^2 x_{0}^{T} (X^TX+ \lambda I_{p})^{-1}(X^TX)^{-1}(X^TX+ \lambda I_{p})^{-1}x_{0}\Big)_{+} \nonumber\\
   &\qquad + \tilde\sigma^{2} x_{0}^{T}(X^TX+ \lambda I_{p})^{-1}X^{T}X (X^TX+ \lambda I_{p})^{-1}x_{0}  \bigg\}. \label{fridge}
\end{align} 
 where  $ \tilde\beta $ and $ \tilde\sigma^2$ are the OLS estimates, and $(\cdot)_{+} = \max \{\cdot,0\}$.
\end{definition}
We can construct a simplified version of fridge by omitting the bias correction
\begin{align}
  \hat\lambda^*_{x_{0},OLS}    
  &= \arg \min_{\lambda} \left\{ (x_{0}^{T}( (X^TX+ \lambda I_{p})^{-1}X^{T}X - I_{p})\tilde\beta)^{2} \right. \\\nonumber
   &\qquad\left. + \tilde\sigma^{2} x_{0}^{T}(X^TX+ \lambda I_{p})^{-1}X^{T}X (X^TX+ \lambda I_{p})^{-1}x_{0}  \right\}. 
\end{align}
which ensures a continuous first derivative. 

In the high-dimensional case, we propose to use ridge regression with the tuning parameter found by CV as the plug-in estimate: 
\begin{definition}[Fridge-ridge]\label{fridgeRidge}
 The fridge-ridge tuning parameter estimate is the minimand of the estimated mean squared error curve
\begin{align*}
  \hat\lambda_{x_{0},ridge} 
   = \arg \min_{\lambda}& \bigg\{ \Big( (\lambda x_{0}^{T} (X^TX+ \lambda I_{p})^{-1}\hat\beta(\hat\lambda_{\cv}))^{2}\\
   - \hat\sigma^{2} \lambda^2 x_{0}^{T} &(X^TX+ \lambda I_{p})^{-1}(X^TX+ \hat\lambda_{\cv}I_{p})^{-1}X^{T}X (X^TX+ \hat\lambda_{\cv} I_{p})^{-1}(X^TX+ \lambda I_{p})^{-1}x_{0}\Big)_{+} \\
   &\qquad + \hat\sigma^{2} x_{0}^{T}(X^TX+ \lambda I_{p})^{-1}X^{T}X (X^TX+ \lambda I_{p})^{-1}x_{0}  \bigg\},
\end{align*} 
 where $\hat\lambda_{\cv}$ is found by cross-validation, giving the standard ridge estimates
$$\hat\beta(\hat\lambda_{\cv}) = (X^TX + \lambda I_{p})^{-1}X^{T}Y, \quad \hat\sigma^2 = \frac{1}{n-df(\hat\lambda_{CV})} \sum_{i=1}^n\left(y_{i}-x_{i}^T\hat \beta(\hat\lambda_{CV})\right)^2,$$
with the effective degrees of freedom $df(\hat\lambda_{CV}) = \tr(X(X^TX +\hat\lambda_{CV})^{-1}X^T)$.
\end{definition}
Other estimators for $\sigma^2$ in the high-dimensional setting have also been proposed \citep{dicker2014variance}.  

For the MSE of the ridge estimate, $\mse\left(\hat\beta(\lambda)\right)$, it has been shown that there always exists a value, $\lambda > 0$, for which the MSE of ridge regression will be smaller than the MSE of OLS \citep{hoerl1970ridge,van2015lecture}. This is also true for the MSE of the prediction $x_{0}^T\beta$; there always exists a tuning parameter value, $\lambda > 0$, with smaller mean squared prediction error. If Eq. \eqref{eq:mse} is rewritten in terms of the singular value decomposition, $X = UDV^T$, as a summation over the singular vectors, $v_{1}, \cdots, v_{p}$, and singular values, $d_{1}, \dots, d_{p}$ 
$$ \mse_{\hat\mu}(\lambda;x_{0}, \beta, \sigma^2) =\left\{ \lambda\sum_{i=1}^p \frac{ x_{0}^{T}v_{i}v_{i}^T\beta}{d_{i}^2+\lambda} \right\}^{2} + \sigma^{2}  \sum_{i=1}^p   \frac{d_{i}^2(x_{0}^{T}v_{i})^2}{(d_{i}^2+\lambda)^{2}},
$$
the first derivative of $\mse_{\hat\mu}$ with respect to $\lambda$ 
\begin{align}
 \frac{\partial }{\partial \lambda}\mse_{\hat\mu}(\lambda;x_{0}, \beta, \sigma^2) &= 2\lambda \left[ \sum_{i=1}^p \frac{ x_{0}^{T}v_{i}v_{i}^T\beta}{d_{i}^2+\lambda} \right]\left[\sum_{i=1}^p \frac{d_{i}^2 x_{0}^{T}v_{i}v_{i}^T\beta}{(d_{i}^2+\lambda)^{2}} \right] - 2 \sigma^{2}  \sum_{i=1}^p   \frac{d_{i}^2(x_{0}^{T}v_{i})^2}{(d_{i}^2+\lambda)^{3}}, \label{derivative}
\end{align}
is always negative in the limit, $\lambda \to 0$:  
 $$\left.\frac{\partial }{\partial \lambda}\mse_{\hat\mu}(\lambda;x_{0}, \beta, \sigma^2) \right|_{\lambda=0}= - 2 \sigma^{2}  \sum_{i=1}^p   \frac{(x_{0}^{T}v_{i})^2}{d_{i}^4}. $$
Thus there always exists a tuning parameter value larger than zero, $\lambda > 0$, for which $\mse(\lambda)$ is smaller than $\mse(0)$. 

The MSE of $\beta^T\beta$ has a single global minimum, which is not the case for the MSE of the prediction, $x_{0}^T\hat\beta(\lambda)$, as the MSE curves in Eq. \eqref{eq:mse} can have several local minima. Extra care therefore needs to be taken when using numerical optimizers to locate the global minimum. There are no explicit solutions for the minima, except for the case of all singular values being equal, and the limit values of the curve are given  
 $$ 
 \lim_{\lambda \to 0}\mse_{\hat\mu}(\lambda;x_{0}, \beta, \sigma^2) =   \sigma^{2} x_{0}^{T}(X^TX)^{-1}x_{0}, \quad \lim_{\lambda \to \infty} \mse_{\hat\mu}(\lambda;x_{0}, \beta, \sigma^2) = (x_{0}^{T}\beta)^{2}.  $$
Thus if $(x_{0}^{T}\beta)^{2} \geq  \sigma^{2} x_{0}^{T}(X^TX)^{-1}x_{0}$, there must exist a global minimum for $\lambda < \infty$. In the reverse case, however, the global minimum can be given in the limit $\lambda \to \infty$. 

\begin{figure}
\centering
\includegraphics[width=0.42\columnwidth,height=0.44\columnwidth,trim = 0 30 340 0, clip]{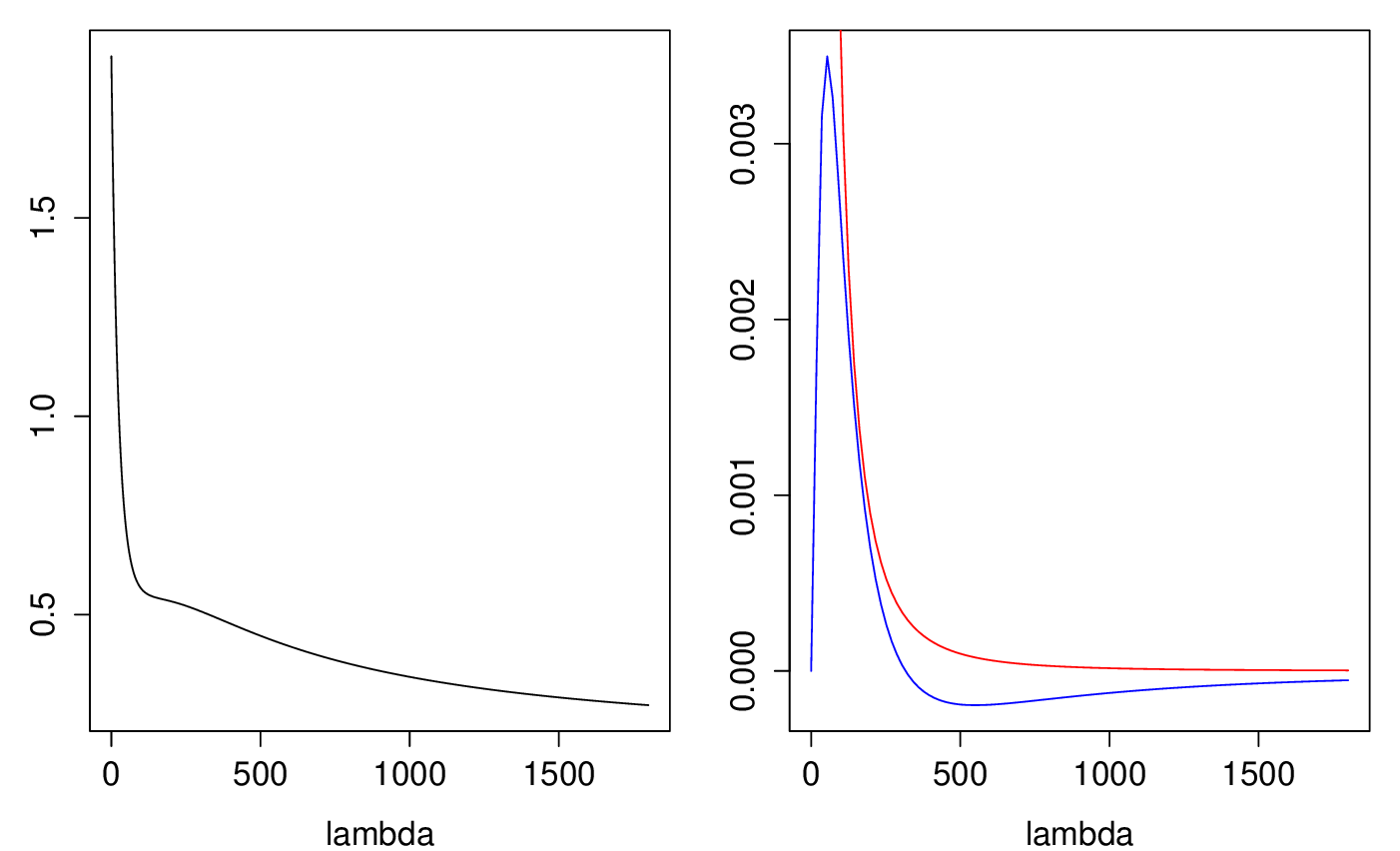}
\includegraphics[width=0.42\columnwidth,height=0.44\columnwidth,trim = 0 30 340 0, clip]{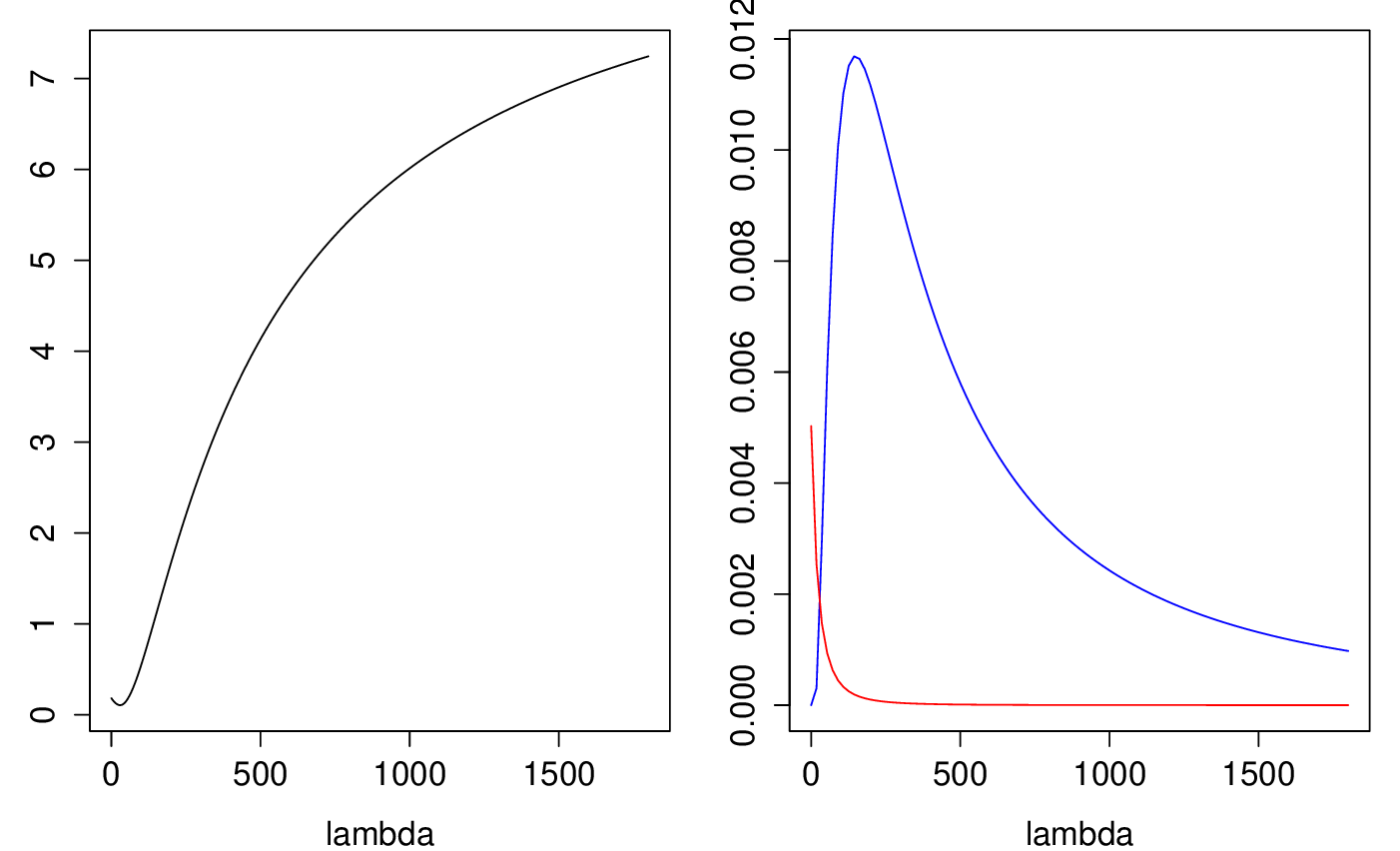}
\caption{MSE curves as a function of $\lambda$ for $p=10$. a) With no critical points the minimum is in the limit $\lambda \to \infty$. b) The classical case with one minimum and a curve increasing towards an asymptote.}
\label{examples}
\end{figure}

\begin{figure}
\centering
\includegraphics[width=0.42\columnwidth,height=0.44\columnwidth,trim = 0 30 340 0, clip]{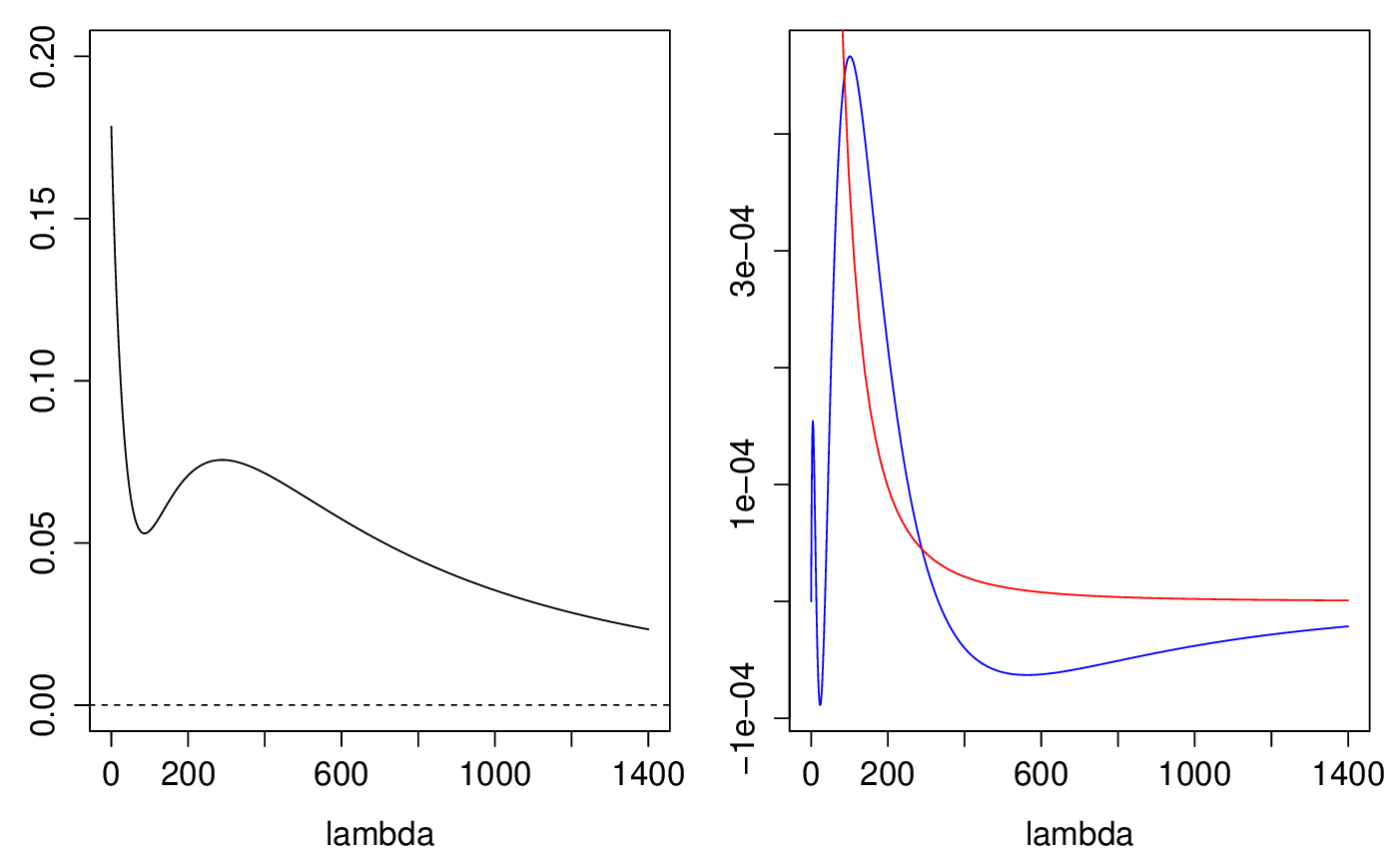}\hspace{0.1cm} 
\includegraphics[width=0.42\columnwidth,height=0.44\columnwidth,trim = 0 30 340 0, clip]{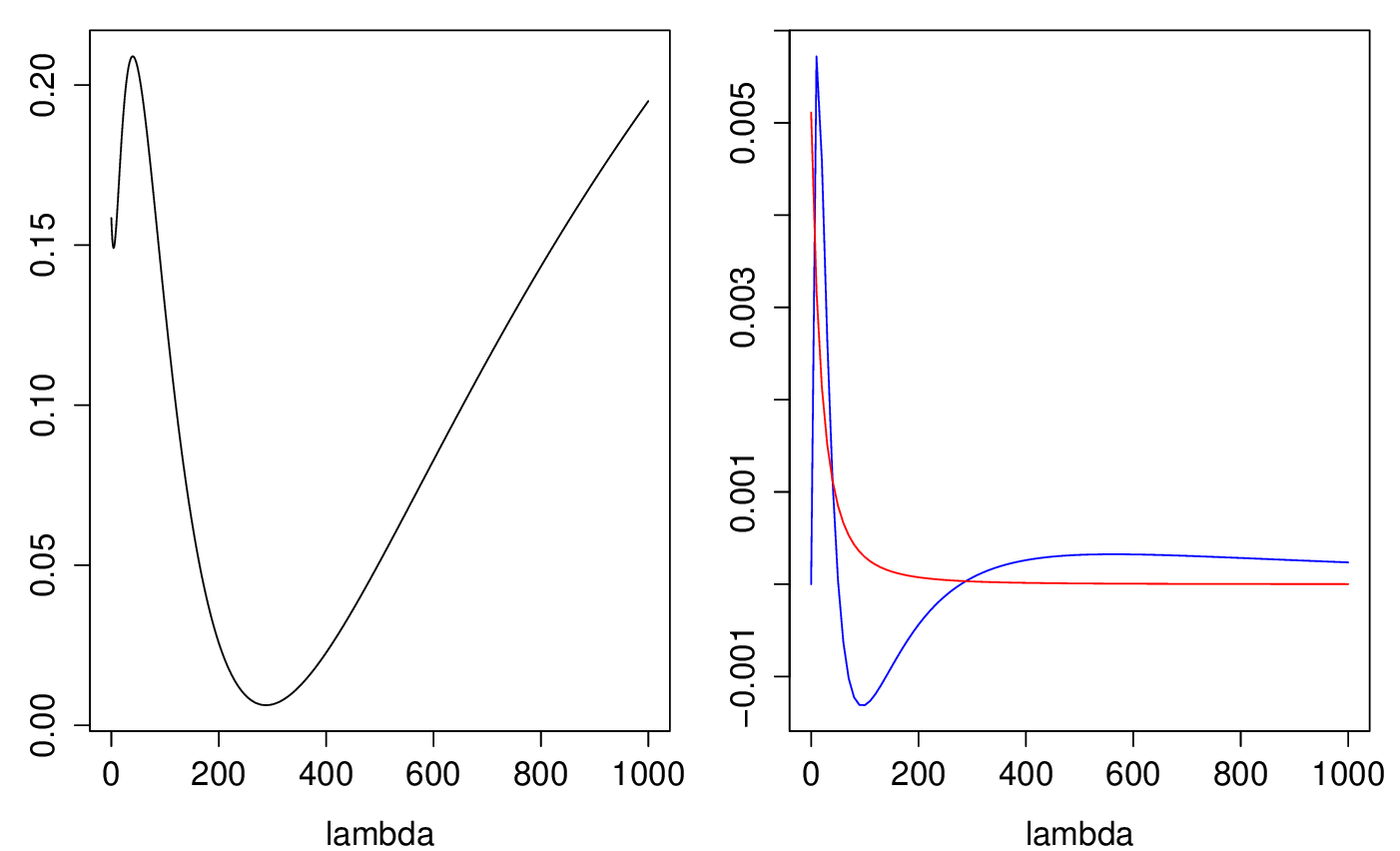}
\caption{MSE curves as a function of $\lambda$ for $p=10$. a) First a minimum and then a maximum, but with an asymptote $x_{0}^T \beta$ \emph{below} the minimum value. b) Two local minima and a local maximum. }\label{examples2}
\end{figure}

We illustrate these characteristics in the case of $p=10, n=100$ with a fixed $x_{0}$ and normally distributed data and regression coefficients, $x_{j} \sim N(0, I_{p}), \beta \sim N(0,I_{p})$. For this setup we typically see zero to three critical points, as shown in Figures \ref{examples} and \ref{examples2}. There can be no critical points (Figure \ref{examples}a), giving the global minimum in the limit, $\lambda \to \infty$ , or one critical point (Figure \ref{examples}), a minimum, giving the global minimum at the local minimum. Further, one can have two critical points (Figure \ref{examples2}a), a minimum and a maximum, or three critical points (Figure \ref{examples2}b), two minima and a maximum, where simulations suggest that the second minimum is always below the first local minimum.

The MSE curves with the plug-in estimates exhibit the same behavior as the oracle curves. As $\lambda \to 0$ the first derivative will always be negative, such that there exists an optimal tuning parameter larger than zero. However, the asymptotic limit of the estimated MSE as $\lambda \to \infty$ changes to 
$$ \lim_{\lambda \to \infty}\widehat \mse(\lambda; \tilde\beta, x_{0},  \tilde\sigma^2)= \max\left\{0, (x_{0}^{T} \tilde\beta)^{2} - \tilde\sigma^{2} x_{0}^{T}(X^TX)^{-1}x_{0}\right\}.$$
The correction of the squared bias by truncating at zero will more often produce a global minimum in the limit $\lambda \to \infty$. 

\section{The benefits of focusing}
The goal of fridge is to lower the expected prediction error of a specific covariate vector $x_{0}$, and to what degree this can be achieved depends on the $x_{0}$, $X$ and $\beta$ in question. In this section, we explore the theoretical characteristics of fridge in simplified examples, to showcase how the relation between $x_{0}$ and $\beta$ affects the procedure. 

We first consider the oracle setting, where $\beta$ and $\sigma^2$ are known, to demonstrate that the effect of focusing acts through the inner product between the vector of covariates and the regression coefficients, $x_{0}^T\beta$. Suppose the covariates are transformed to give a diagonal covariance matrix with equal entries,  
$$X^TX = \diag(M, \dots, M) = M I, \quad M = \sum_{i=1}^n x_{i,j}^2, $$
 such that the columns of $X$ are orthogonal. This is an artificial data matrix allowing for an explicit solution with the crucial aspect being equal diagonal entries. The oracle MSE from Eq. \eqref{eq:mse} can then be written out with an explicit solution
$$\mse_{\hat \mu}(\lambda; x_{0},\beta, \sigma^2) 
   = (x_{0}^{T}\beta)^2 \frac{\lambda^2}{(M+ \lambda)^2}+ \sigma^{2} x_{0}^{T}x_{0}\frac{M}{(M+ \lambda)^2}, \quad \lambda_{x_{0}} =   \frac{\sigma^2x_{0}^{T}x_{0}}{(x_{0}^{T}\beta)^2}. 
$$
The oracle tuning parameter is thus controlled by $x_{0}^{T}\beta$, the inner product between the focus covariate vector and regression coefficients, and the tuning parameter can be re-expressed in terms of the geometry of $x_{0}$ and $\beta$, specifically the length of $\beta$, $\|\beta\|$, and the angle between the vectors $x_{0}$ and $\beta$, $\alpha_{x_{0}}$:
\begin{equation}
  \lambda_{x_{0}} =    \frac{\sigma^2 }{\|\beta\|^2 \cos^2 \alpha_{x_{0}}}, \label{orthotuning}
\end{equation} It is evident that the length of $x_{0}$, $\|x_{0}\|$, does not influence the value of the oracle tuning in the orthogonal case. In addition, if $p=1$, the $x_{0}$ in Eq. \eqref{orthotuning} cancels out, such that $\lambda_{x_{0}}$ does not depend on $x_{0}$ at all and the estimator has in some sense \emph{lost} its focus. 

The covariate vector $x_{0}$ influences the oracle tuning through its relation to $\beta$ as the angle $\alpha$ measures how close the prediction is to the mean response. When the true outcome is close to the mean of $Y$, $\cos \alpha_{x_{0}}$ will be close to zero (meaning $x_{0}$ and $\beta$ are close to being orthogonal). This causes the oracle tuning parameter to blow up, $\lambda \to \infty$, shrinking the estimated prediction towards the mean. The length of $\beta$ on the other hand acts as a measure of the signal strength, such that larger values of $\beta$, i.e.~a stronger signal, warrants a stronger penalization in the optimal case, while weaker signal requires less penalization. As previously stated, the resulting prediction error of fridge in the oracle case will be uniformly smaller than the OLS prediction error, corresponding to $\lambda = 0$;
  $$ \mse_{\hat \mu}(\lambda_{0}; x_{0},\beta, \sigma^2) = \frac{\sigma^2 x_{0}^Tx_{0} (x_{0}^T\beta)^2}{\sigma^2 x_{0}^Tx_{0}+(x_{0}^T\beta)^2}< \mse_{\hat \mu}(0; x_{0},\beta, \sigma^2) =\sigma^2 x_{0}^Tx_{0}.$$
     
The effect of the data matrix $X$ is best understood as a modification of $x_{0}$ and $\beta$, relative to the orthogonal case. Consider the general case where the singular value decomposition of the data matrix is $X = U D V^T$, giving the mean square error
\begin{equation}
   \mse_{\hat\mu}(\lambda; x_{0},\beta,  \sigma^2)=\left\{  \lambda x_{0}^{T} V (D^2+ \lambda I_{p})^{-1} V^T \beta \right\}^{2} + \sigma^{2}  x_{0}^{T}V D (D^2+ \lambda I_{p})^{-1} DV^Tx_{0}, 
\end{equation}
where the matrix of singular vectors $V$ rotate the original $x_{0}$ and $\beta$. The data matrix determines the value of $\lambda$ by projecting the $x_{0}$ and $\beta$ along the singular vectors and up-weighting the vectors associated with large singular values. The data matrix, therefore, gives the premise for which directions in the covariate space that are considered more important. Hence, how $x_{0}$ and $\beta$ are spanned by the first singular vectors of $X$ will together determine the optimal value of the tuning parameter. If all singular values are equal, such that all directions carry the same weights, the data matrix $X = U (M I_{p})V^T$ works as a rotation matrix through the singular vectors, and $V^Tx_{0}$ and $V^T\beta$ can be viewed as \emph{new} a covariate vector and regression coefficients oriented along the singular vectors: 
$$ \mse_{\hat \mu}(\lambda; x_{0},\beta, \sigma^2) = (x_{0}^{T}V V^{T}\beta)^2 \frac{\lambda^2}{(M+ \lambda)^2}+ \sigma^{2} x_{0}^{T}VV^{T}x_{0}\frac{M}{(M+ \lambda)^2}, \quad \lambda_{x_{0}} =   \frac{\sigma^2x_{0}^{T}V V^{T}x_{0}}{(x_{0}^{T}V V^{T}\beta)^2}.$$
%

If instead some of the singular values are substantially larger than the rest, the first singular vectors of the data matrix will dominate, and it is crucial how $x_{0}$ and $\beta$ are spanned by the first singular vectors. If all variables are equally correlated
$$ X^TX = \begin{bmatrix}
              M & R & \hdots & R \\
   R & M &  & R \\
   \vdots &  &  \ddots & \vdots \\
   R & R & R & M
             \end{bmatrix},   $$
the first singular vector, $v_{1} = [1, 1, \dots, 1]^T$, has a singular value substantially larger than the rest: $d_{1} \gg d_{2} = \cdots = d_{p}$. For such a data matrix, the part of $x_{0}$ and $\beta$ spanned by $v_{1}$ will be heavily emphasized. If $\beta$ is spanned by $v_{1}$ alone and $x_{0}$ is orthogonal to $v_{1}$, the bias part becomes zero, forcing $\lambda \to \infty$ and the prediction towards the mean. In the reverse situation, if $\beta$ is not properly spanned by $v_{1}$, the same shrinkage towards the mean will occur. But when $\beta$ and $x_{0}$ are spanned by $v_{1}$ alone, the tuning parameter will approximately be given by the length of $\beta$, scaled by the largest singular value, $\lambda_{x_{0}}= \frac{\sigma^{2}}{d_{1}^2\|\beta\|^2}.$ 

\begin{figure}
\centering
 \includegraphics[width=0.51\columnwidth]{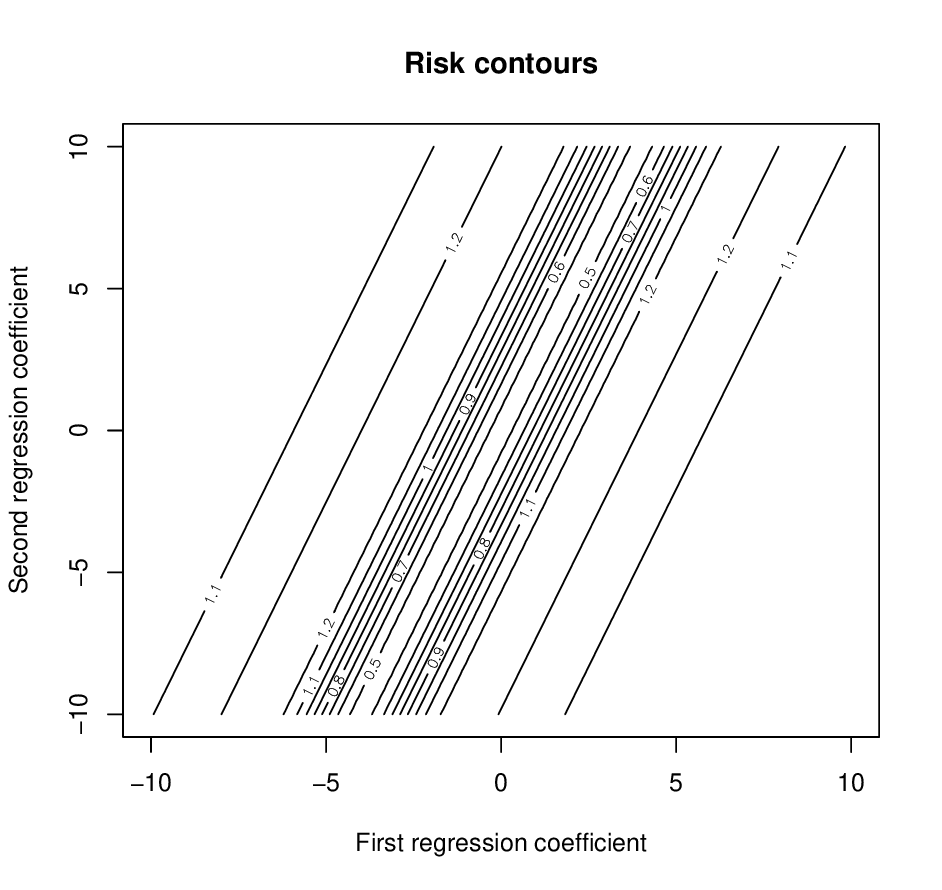} \vspace{0.2cm}
 \includegraphics[width=0.48\columnwidth]{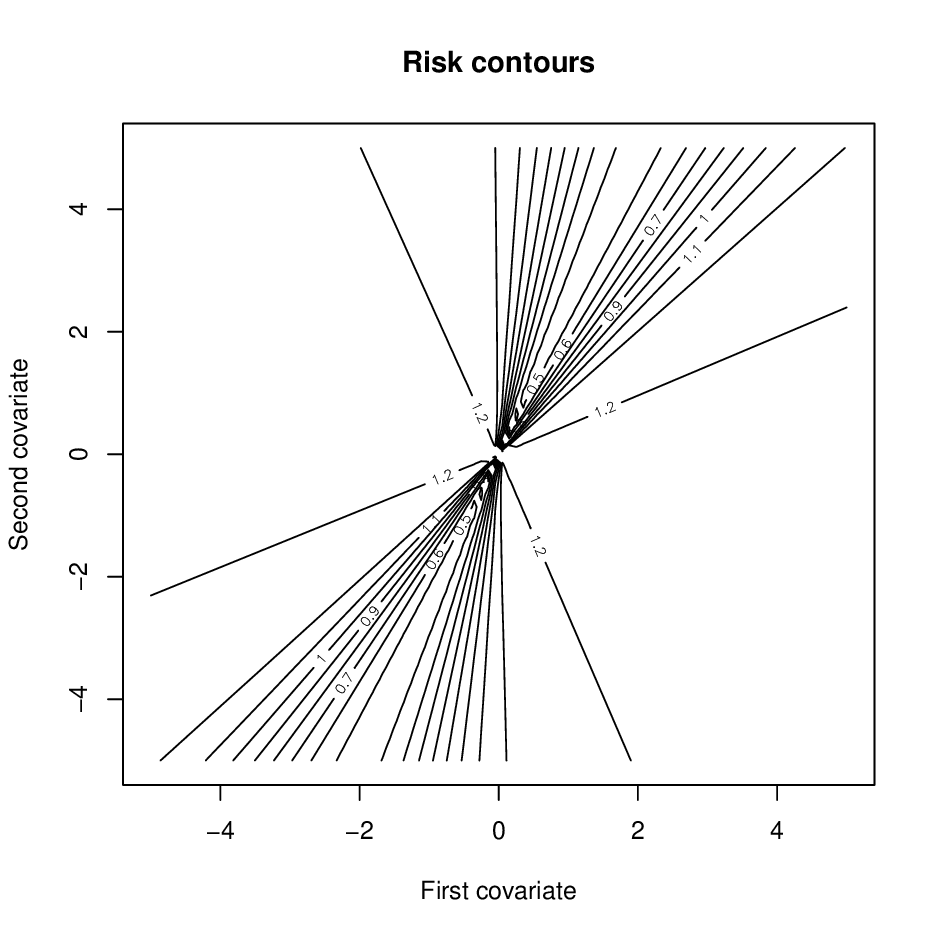}
  \caption{ Contour plots of the risk of fridge-OLS for an orthogonal data matrix scaled relative to the OLS risk a) as a function of $\beta_{1}$ and $\beta_{2}$ for a fixed covariate vector $x_{0} = (-5,2)$, and b) as a function of $x_{0,1}$ and $x_{0,2}$ for fixed regression coefficients $\beta = (-5,2)$. }\label{riskSurface}
\end{figure}

It is also a question how much of the oracle optimality is lost by estimating the tuning parameter through the plug-in approach. In the orthogonal case, the distribution of the estimated bias of fridge-OLS is given 
$$ \widehat \bias = - \frac{\lambda}{M+\lambda} x_{0}^T\tilde\beta, \quad \var \widehat \bias = \frac{\tilde\sigma^2 \lambda^2 x_{0}^Tx_{0}}{M(M+\lambda)^2},$$
such that the estimated fridge-OLS tuning parameter is explicitly given 
$$\hat \lambda_{x_{0},OLS} = \frac{\tilde\sigma^2 Mx_{0}^{T}x_{0}}{\left(M (x_{0}^T\tilde\beta)^2 - \tilde\sigma^2x_{0}^Tx_{0}\right)_{+}}.$$ 
When combining the estimated tuning parameter with the ridge prediction
$$x_{0}^T\hat \beta (\hat\lambda_{x_{0},OLS}) = 
\begin{cases}
 0 & \text{if }  |x_{0}^T\tilde\beta| \leq\tilde\sigma \sqrt{x_{0}^Tx_{0}/M} ,\\
\displaystyle{\frac{(x_{0}^T\tilde\beta)^2 - \tilde\sigma^2x_{0}^Tx_{0}/M}{ (x_{0}^T\tilde\beta)^2}\;x_{0}^T\tilde\beta} & \text{if } |x_{0}^T\tilde\beta| > \tilde\sigma \sqrt{x_{0}^Tx_{0}/M},
\end{cases}$$
the risk of fridge-OLS is given
\begin{align*}
 \risk \left( x_{0}^T\hat \beta (\hat\lambda_{x_{0}, OLS}) \right) 
 &=\begin{cases}
 (x_{0}^T\beta)^2 & \text{if } |x_{0}^T\tilde\beta| \leq\tilde\sigma \sqrt{x_{0}^Tx_{0}/M},\\
\E \left(\frac{(x_{0}^T\tilde\beta)^2 - \tilde\sigma^2x_{0}^Tx_{0}/M}{ (x_{0}^T\tilde\beta)^2}\;x_{0}^T\tilde\beta  - x_{0}^T\tilde\beta\right)^2 & \text{if } |x_{0}^T\tilde\beta| > \tilde\sigma \sqrt{x_{0}^Tx_{0}/M}.
\end{cases}
\end{align*}
If the residuals are assumed to be normally distributed, $\varepsilon_i \sim N(0,\sigma^2)$, the risk can be visualized in two dimensions, for either fixed $x_{0}$ or $\beta$. Figure \ref{riskSurface} displays contour plots of the risk of fridge-OLS for an orthogonal data matrix, scaled to give OLS risk equal to 1. The left panel shows a contour plot of the risk as a function of $\beta$ for fixed covariates, $x_{0}=[-5,2]$, illustrating that the risk of fridge-OLS is constant along lines parallel to $x_{0}^T\beta =0$. Thus fridge-OLS will have lower risk than OLS within a trench following the line $x_{0}^T\beta =0$. The right panel shows the risk as a function of $x_{0}$ for fixed regression coefficients, $\beta=[-5,2]$, and then the risk will be constant along radial lines from the origin. This illustrates that it is the angle between $x_{0}$ and $\beta$, and not the length of $x_{0}$, which determines the benefit of the focused approach. The fridge-OLS will have lower risk than OLS within a \emph{cone}, oriented along the line $x_{0}^T\beta =0$. 

\section{Fridge for logistic ridge regression}
The focused approach to ridge regression can also be extended to generalized linear models by utilizing parametric bootstrap to obtain expressions for the variance and squared bias. Consider logistic ridge regression: independent responses, $y_{i} \in \{0,1\}$ for $i=1, \dots, n$, are distributed as
$$ y_{i} \sim \text{Bernoulli}(\text{logit}^{-1}(x_{i}^T\beta)),$$
for covariate vector $x_{i}$ with the link function $\text{logit}^{-1}(x_{i}^T\beta) = \exp(x_{i}^T\beta)/(1 +  \exp(x_{i}^T\beta))$, where the regression coefficients, $\hat \beta$, are estimated by maximizing the penalized log-likelihood with an $L_{2}$ penalty 
$$ \sum_{i=1}^n [y_{i} \log(p_{i}) + (1- Y_{i})\log(1-p_{i})]- \lambda \sum_{j=1}^p \beta_{j}^2, \quad p_{i} = \text{logit}^{-1}(x_{i}^T\beta), $$
typically using the Newton-Raphson algorithm \citep{le1992ridge}. 

We then propose the following procedure for extending the fridge concept to logistic regression: 
\begin{enumerate}
 \item Use parametric bootstrap with plug-in $\tilde \beta$ to simulate $r=1,\dots,M$ bootstrap samples of $n$ observations $Y^{(r)}$
 $$ y_{i}^{(r)} \sim \text{Bernoulli}(\text{logit}^{-1}(x_{i}^T\tilde\beta)), \qquad i = 1, \dots, n.$$
 \item Over a suitable grid of $\lambda$, holding the tuning parameter value fixed: 
 \begin{itemize}
  \item calculate $\hat\beta_{\lambda}^{(r)}$ for each bootstrap sample,
  \item find the squared bias and variance of $x_{0}^T\hat\beta_{\lambda}^{(r)}$, compared to $x_{0}^T\tilde\beta$ as the population parameter,
  \item add the terms, yielding the MSE. 
 \end{itemize}
 \item Set the estimate $\hat \lambda_{x_{0}}$ to the tuning parameter value with the smallest MSE over the grid of $\lambda$.   
\end{enumerate}
With the use of the \verb|glmnet| \verb|R| package, the estimate $\hat\beta_{\lambda}^{(r)}$ can be calculated for a long sequence of $\lambda$ within each bootstrap iteration, greatly increasing calculation speed. Similar fridge procedures of Cox's proportional hazard regression may also be established.   

\section{Simulation: comparison with cross-validation} \label{sec:simulations}
To compare the predictive performance of fridge and ridge with CV, we perform simulations with both simulated and real data. $K$-fold cross-validation is the most widely used fine-tuning procedure, probably due to its conceptual simplicity: The data is divided into $K$ folds with each part held out and predicted by fitting a model on the remaining folds. A range of tuning parameters can then be tested and one will choose the value with the lowest error, averaged over all folds \citep{stone1974cross, allen1974relationship}. Currently $10$- or $5$-fold cross-validation has become the default approach in modern statistics and machine learning \citep{hastie2001elements}. In the case of ridge regression, leave-one-out cross-validation (LOOCV) error has an explicit expression \citep{golub1979generalized}, thus being particularly easy to calculate
$$\hat\lambda_{\cv} = \arg\min_{\lambda}\sum_{i=1}^n \left( \frac{y_{i} - x_{i}^T\hat{\beta}(\lambda)}{1-w_{ii}} \right)^2, \quad w_{ii} = x_{i}^T(X^T X+\lambda I)^{-1}x_{i},$$ and LOOCV will be used for the remainder of the paper. 

\begin{figure}
\centering
\includegraphics[width=\columnwidth]{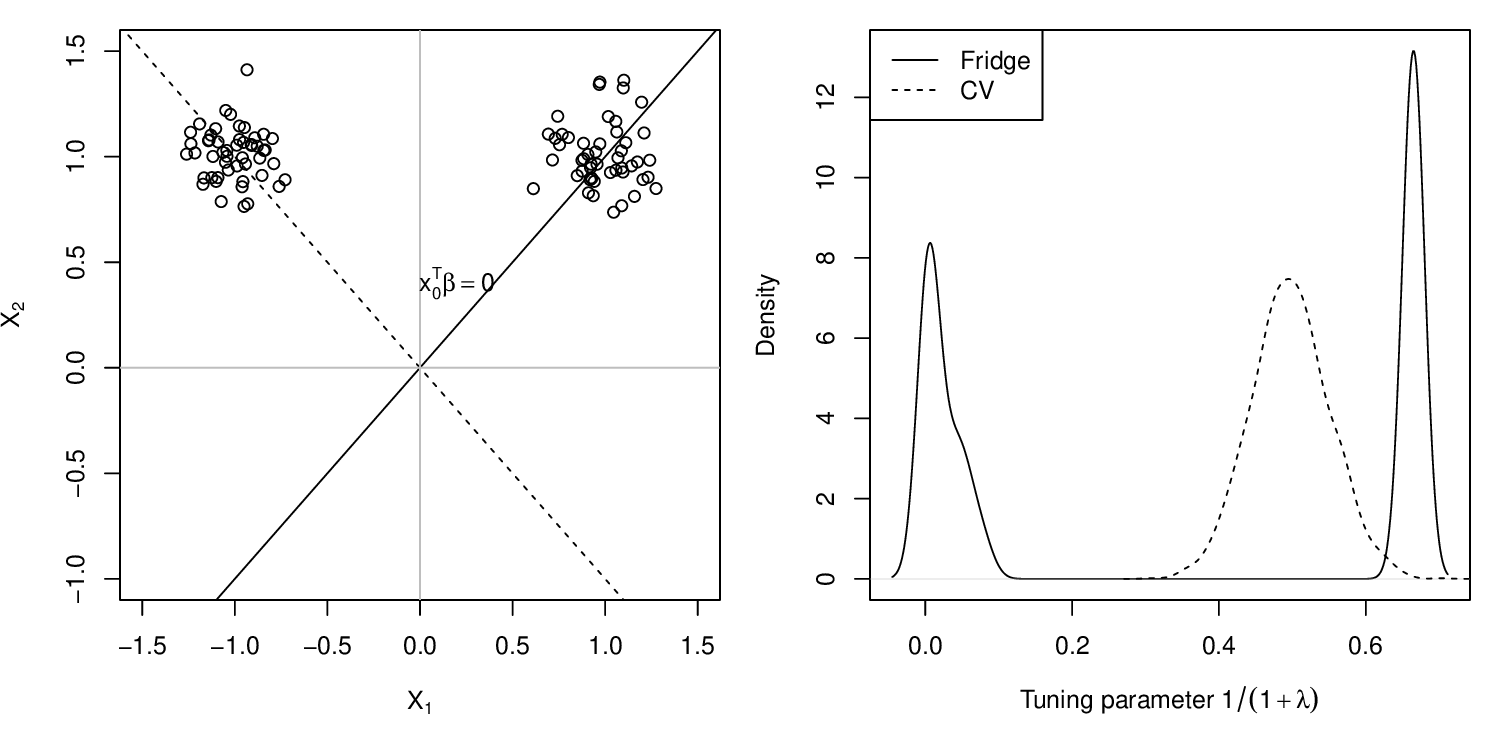}
\caption{a) Data concentrated in two distinct clusters centered at $(1,1)$ and $(-1,1)$ b) Density plot of $1/(\lambda+1)$ for fridge (black) and cross-validation (dashed) for $\beta = [-1,1]$ and $\sigma^2 = 1$. 
}\label{lambdaDist}
\end{figure}

To illustrate the difference between the focused tuning and cross-validation, consider an example where the data matrix consists of different clusters. Figure \ref{lambdaDist}a) shows a data example $(p=2)$ with two distinct clusters centered at $(1,1)$ and $(-1,1)$, respectively. If the regression coefficients are given $\beta = [-1,1]$, the outcome for the right cluster will be close to zero, $x_{i}^T\beta \simeq 0$, while the outcome for the left cluster will be close to two, $x_{i}^T\beta \simeq 2$. The line implied by $x^{T}\beta = 0$ is marked in black. The clusters will then require a very different level of penalization to produce an optimal prediction; the right cluster requires a stronger penalization and the left cluster requires a weaker penalty. Figure \ref{lambdaDist}b) shows the distribution of the covariate vector specific tuning, $1/(1+\lambda_{x_{0}})$ in red, for each of the observations seen in Figure \ref{lambdaDist}a). The corresponding distribution of the tuning parameter estimated by leave-one-out cross-validation, $1/(1+\hat\lambda_{CV})$ over multiple sets of simulated $y_{i}$ is shown in black. Figure \ref{lambdaDist}b) displays clearly that the difference in optimal tuning parameter for the two clusters are captured by the fridge procedure, with the right cluster corresponding to a large tuning parameter value and the left cluster to a small tuning parameter value. Cross-validation, on the other hand, estimates an overall tuning parameter, averaging over all individuals, and thus selects a tuning parameter value inappropriate for both  clusters. 

\begin{figure}[!h]
\centering
\includegraphics[width=0.6\columnwidth]{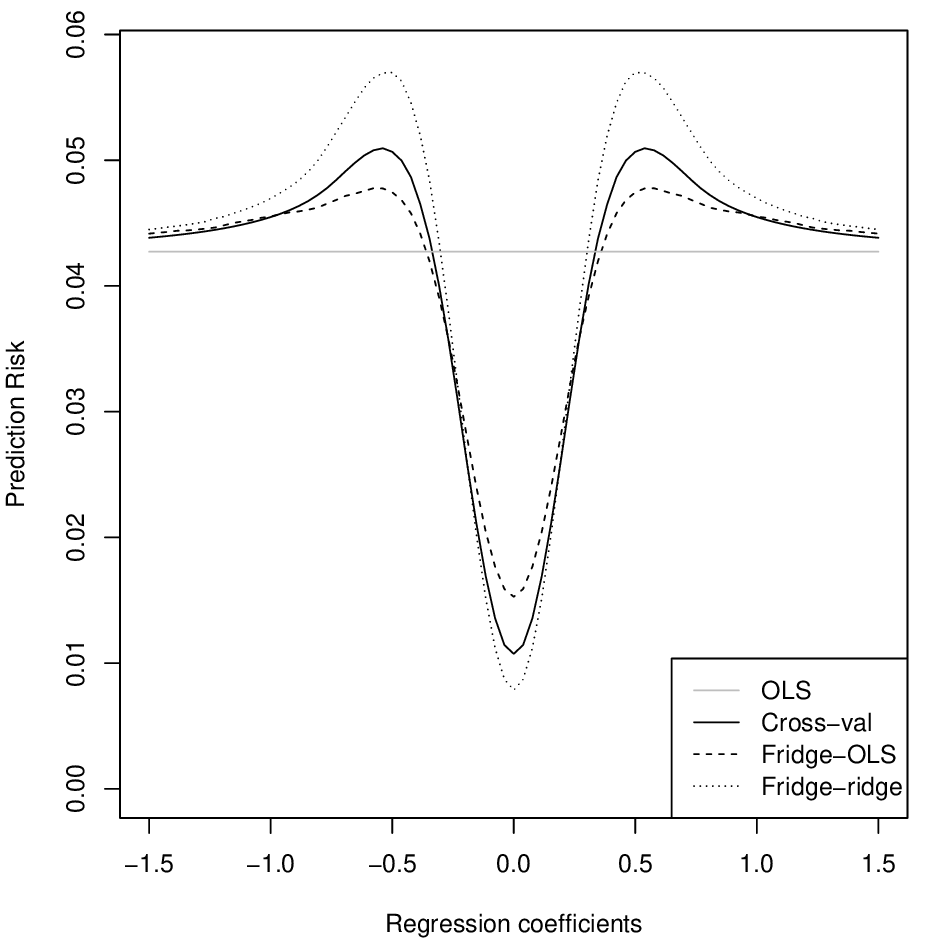}
 \caption{Risk functions for OLS, ridge with cross-validation, fridge-OLS, fridge-ridge for varying $\beta$.}
 \label{riskUniform}
\end{figure}

In low dimensions, the risk of fridge-OLS and fridge-ridge can be compared to ridge with cross-validation for varying $\beta$. In Figure \ref{riskUniform}, the average squared prediction error is shown in the case of $p=2 $ and $ n=50$, when $\beta = [b,b]$ and the data matrix $X$ and focus $x_{0}$ are drawn from a uniform distribution $X, x_{0} \sim U(-1,1)$. It is seen that the risk of fridge-OLS is slightly higher than cross-validation for $\beta$ close to zero, but lower for medium $\beta$. The fridge-ridge has the same risk as ridge with cross-validation for $\beta$ close to zero, but a higher risk for medium $\beta$.

\begin{figure}
\centering
\includegraphics[width=0.6\columnwidth]{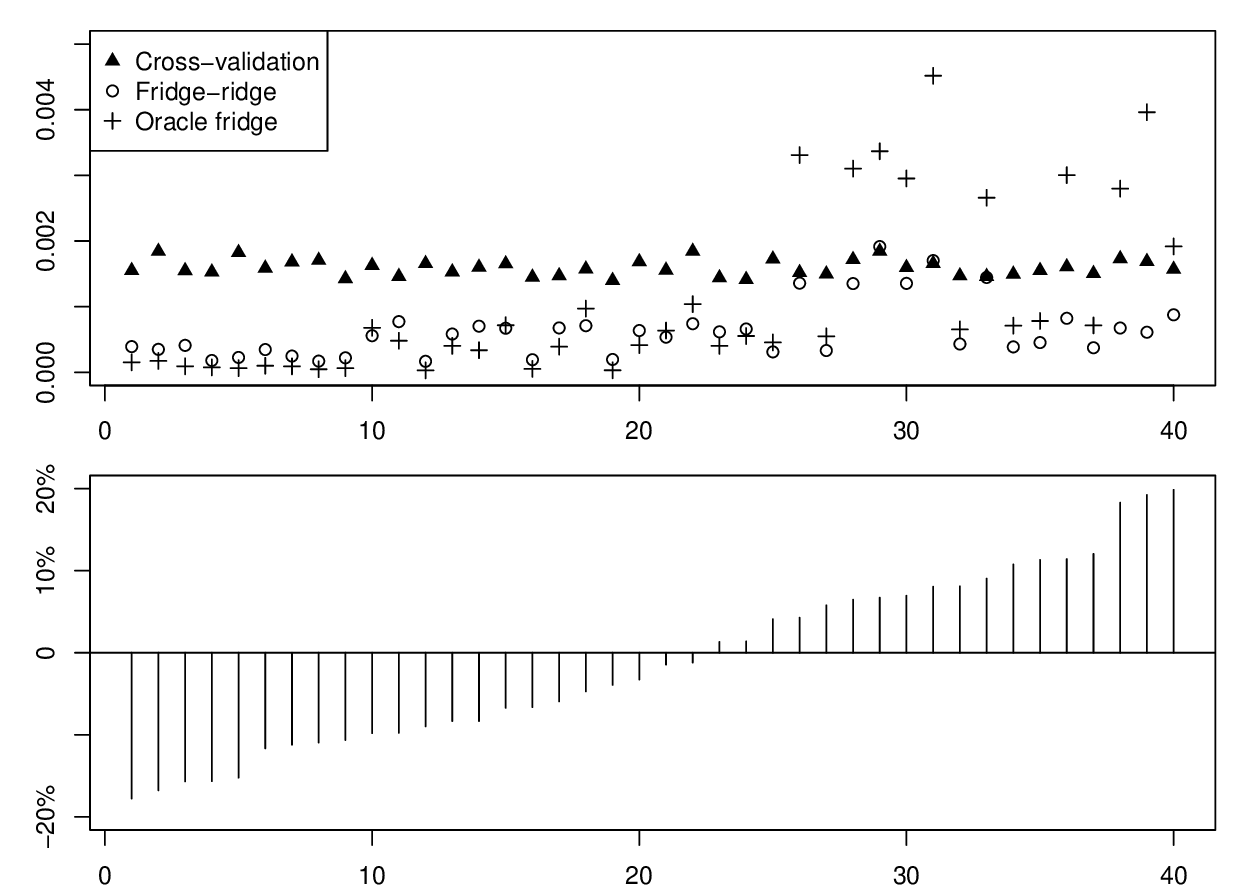}
 \caption{Top panel: the inverse of the estimated tuning parameter for each out-of-sample observation given by cross-validation (triangle), oracle fridge (cross) and fridge-ridge (circle). Bottom panel: the corresponding relative difference in MSE for fridge-ridge compared to ridge with cross-validation. 
}\label{simulation}
\end{figure}

Lastly, we compare fridge to standard ridge in a simulation study based on real high-dimensional covariates, with simulated $y_{i}$ and known $\beta$ coefficients. The data consist of gene expression profiles measured in 40 glioma patients \citep{moeckel2014response}, and the data are described and analyzed in detail in Section \ref{sec:examples}. The simulation was done on the 1000 genes with the largest variance, and was carried out by drawing standard normally distributed residuals, $\varepsilon_i \sim N(0,1)$, and regression coefficients from the normal distribution, $\beta \sim N(0,0.05)$, to ensure a suitable signal-to-noise ratio. We then simulated 200 sets of outcomes $Y$ and calculated the average squared prediction error of fridge-ridge and ridge regression with cross-validation for each individual. The bottom panel of Figure \ref{simulation} shows the relative difference in the MSE of fridge-ridge compared to ridge with cross-validation, ordered from best to worst, and fridge gives a lower MSE for a majority of the observations. The upper panel shows the out-of-sample estimated tuning parameter on an inverse scale for CV, fridge-ridge and the oracle fridge, and demonstrates that fridge-ridge estimates well the large oracle tuning parameters, but is less precise when estimating the smallest oracle values. 

\section{Data examples}\label{sec:examples}
We demonstrate the proposed fridge-ridge procure in two examples where accurate prediction of individual patients can be considered more important than overall accuracy; specifically the prediction of complication risk and the response of treatment. 

\subsection{Prediction of weight gain}
We demonstrate the fridge procedure in a study investigating whether gene expression can be used to predict weight gain \citep{cashion2013expression}, available in the EMBL-EBI ArrayExpress database (www.ebi.ac.uk/arrayexpress) under accession number E-GEOD-33070. Kidney transplant recipients are known to gain substantial weight during the first year after transplantation, with a reported average increase of 12 kg \citep{patel1998effect}. Such large weight gain over a short time period results in increased risk for adverse health effects, e.g. cardiovascular disease, contributing to an overall worse outcome for patients. The weight gain has been partly explained by patients experiencing better appetite, due to the use of prescribed steroids, and by patients having less restrictive diets after transplantation. However, steroid-free protocols do not alone reduce the risk of obesity, suggesting that other causes also contribute \citep{elster2008obesity}. 

Weight gain is fundamentally caused by a too high intake of calories, relative to the energy expenditure, but the individual response is seen to be substantial. Hence genetic variations have been thought to be a contributing factor, and several genes have already been linked to obesity and weight gain. To investigate the predictive power of genomic data regarding  weight gain, gene expression profiles were measured in adipose tissue taken from kidney transplant patients \citep{cashion2013expression}. Subcutaneous adipose tissue was considered particularly well-suited as it is involved in appetite regulation and can be easily obtained from the patients during surgery. Tissue samples from 25 transplant patient were collected at the time of surgery, and mRNA levels were measured using Affymetrix Human Gene 1.0 ST arrays, obtaining gene expression profiles for 28 869 genes. Patients were weighed at transplantation and at a follow-up time of one year, resulting in a one-year recorded weight gain.  Additional covariates, such as race and gender, were collected to adjust for possible confounding, but gene expression variability was not associated with neither characteristics. 

As excessive weight gain can have severe consequences for the patients, the goal is to predict the future weight increase based on the available gene expression profiles. When a large increase in weight is predicted, additional measures such as diet restrictions or physiotherapy could be set in effect. It is thus important to predict the weight gain of each transplant recipient as accurate as possible. In such a setting, a focused tuning parameter tailored to the covariate vector of each patient could give an advantage. 

We will predict each of the 25 measured patients out-of-sample based on the remaining 24 observations, considering the hold-out observation as the new covariate vector $x_{0}$.
The plug-in estimates for $\tilde\beta$ and $\tilde \sigma^2$ are given by ridge regression with the LOOCV tuning parameter, following Definition \ref{fridgeRidge}. 

\begin{figure}
\centering
\includegraphics[width=0.6\columnwidth]{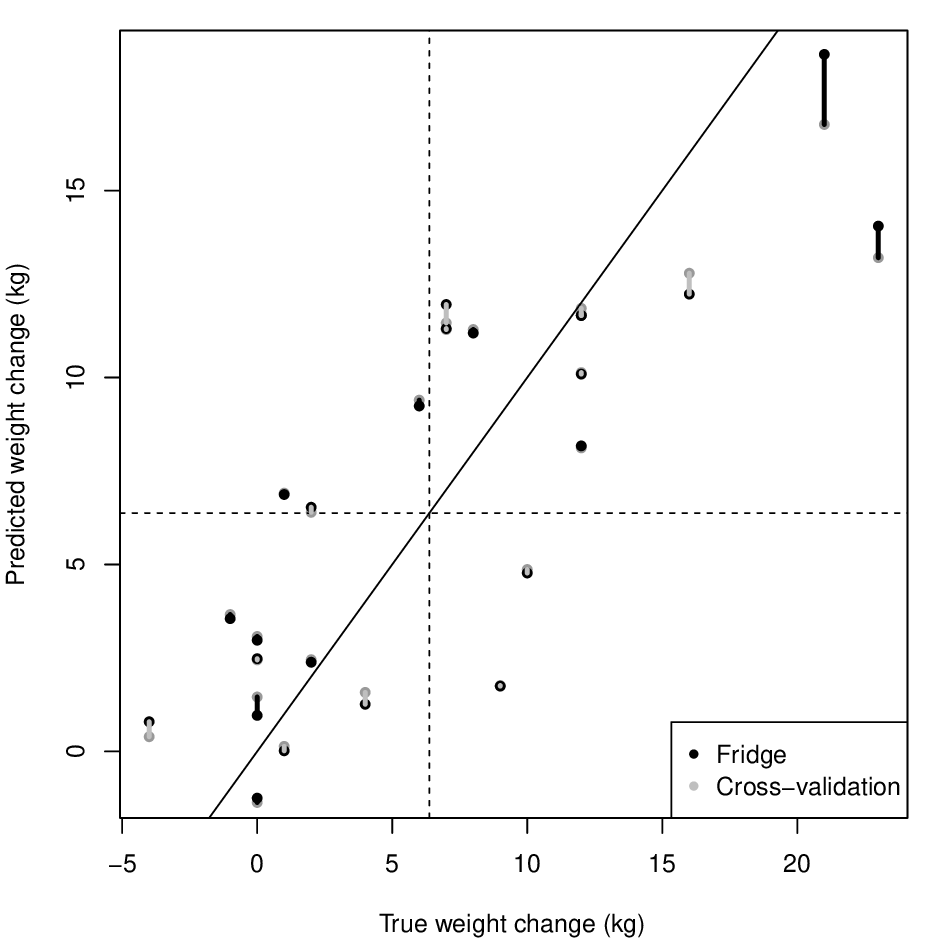}
 \caption{The out-of-sample predicted change in weight for fridge (in black) and ridge with cross-validation (in gray) against the true change in weight. The difference between the fridge and ridge with CV is colored according to the method with the lowest squared prediction error.  }\label{adipose}
\end{figure}

Figure \ref{adipose} shows the out-of-sample predicted change in weight estimated by fridge (in black) and ridge with cross-validation (in gray), plotted against the true change in weight. The difference between the predictions from the two methods is colored black when fridge gives the lowest error, and grey otherwise. In general, it is seen that both methods achieve a good out-of-sample predictive performance. The observations in the upper right corner of Figure \ref{adipose} are penalized less by the focused approach, compared to cross-validation, which results in smaller prediction error. This demonstrates how fridge aims to give a smaller penalty to predictions far from the outcome mean and a larger penalty to the predictions close to the outcome mean.    

Fridge gives a smaller prediction error compared to ridge with cross-validation for around 44\% of the observations. But the overall improvement in the individual predictions gives an average squared prediction error of 15.60 for fridge and 16.24 for standard ridge, which yields a 4.0 \% decrease when using the focused approach. The average parameter estimate over the leave-one-out models was, $\hat \sigma^2 = 6.7$, for the variance and, $\hat \lambda_{cv} = 12.2$, for the cross-validation tuning parameter.   

\subsection{Prediction of treatment response}
The prediction of treatment response, in particular within cancer treatment, is an important application in the field of personalized medicine. We illustrate the logistic fridge procedure with treatment response data in glioma tumor samples.

High-grade gliomas, cancer tumors in the brain, are amongst the most deadly human tumors, and in treatment trials only around 20\% of patients respond to therapy. It has been investigated whether gene expression can be used to identify glioma patients that will profit from cancer therapy \citep{moeckel2014response}, and the data are publicly available in the EMBL-EBI ArrayExpress database (www.ebi.ac.uk/arrayexpress) under accession number E-GEOD-76900. Decisions regarding treatment for glioma patients are currently based on age and performance status \citep{weller2012individualized}, and increased use of molecular markers would be beneficial. As a follow-up validation study, 18 samples of high-grade gliomas were treated with Sunitinib (tyrosine kinase inhibitor) to measure treatment response in terms of decrease in proliferation rate, or cell growth, over a 6 hour time period, compared to a control sample of the same tissue \citep{moeckel2014response}. To illustrate logistic fridge, we dichotomize the outcome in a high and low treatment response group, relative to the median value. Before treatment, genetic expression profiles of 19 410 genes were measured using Affymetrix Human gene 1.1 arrays, and for analysis, we used the 3000 genes most correlated with the treatment response. 

\begin{figure}
\centering
\includegraphics[width=0.6\columnwidth]{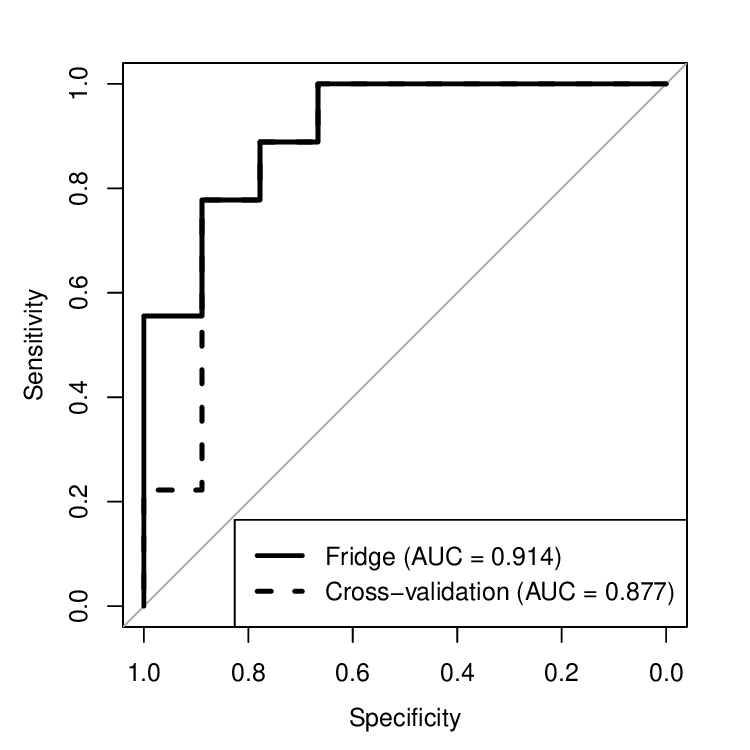}
 \caption{The ROC curve for fridge-ridge (black) and ridge with cross-validation (dashed).}\label{glioma}
\end{figure}

Figure \ref{glioma} shows the ROC curves for the logistic out-of-sample prediction of the 18 samples, with the predictions based on fridge (black line) and the ridge with cross-validation (dashed line). It is seen that the ROC curves are identical for most of the predictions probably due to the low number of samples, but the fridge preforms better in predicting the zero values. Fridge performs better in terms of area under the curve (AUC) with an AUC value of 0.914, compared to an AUC value of 0.877 for ridge regression with CV. 

\section{Discussion}
The development of personalized medicine will increase the demand for new prediction methodologies targeting the individual, where one possible approach is to allow tuning parameter(s) to vary with the covariate values for which the prediction is to be given. A covariate vector specific tuning parameter, $\lambda_{x_{0}}$, can be defined as the minimizer of the expected prediction error, and we have proposed to estimate this tuning parameter by plugging in separately obtain regression coefficients and noise variance estimates in the theoretical MSE expressions. As the MSE approach minimizes the error on a \emph{population level}, it allows for the added focus on specific individuals. Standard cross-validation, on the other hand, cannot be focused in the same way, as it relies on averaging over all observed outcomes to minimize the \emph{sample} prediction error. Where fridge tries to minimize the expected squared prediction error for each individual, cross-validation minimizes the observed squared prediction error over all  individuals. The difference lies in averaging over the theoretical distribution of $y_{i}$, instead of averaging over the observed outcomes. Our simulations demonstrate, however, the adaptability and robustness of cross-validation, and that the benefits of focusing can be lost if the utilized plug-in estimates are not precise enough. Future works should explore whether one can predetermine for which covariate values the fridge approach will out-perform CV. Further, also alternative loss functions and more direct approaches to estimation should be explored. 

The use of ridge regression combined with cross-validation as the plug-in estimate can also be set in the following conceptual framework: in an initial step, cross-validation is used to establish an average or overall suitable level of penalization, followed by a focusing step where the average value is tweaked to a necessary stronger or weaker penalization, depending on the individual predictions. The focused tuning parameter is also connected to random effect models when viewed in the Bayesian context. As ridge regression corresponds to a Gaussian prior on the regression coefficients with a variance inversely proportional to the tuning parameter, a covariate vector specific $\lambda_{x_{i}}$ can be formulated as a Gaussian prior with a variance specific to the covariate vector. This can further be viewed as an individual-specific scaling of $\beta$:
$$ y_{i} =  x_{i}^T \lambda_{x_{i}}^{-1/2}\beta + \varepsilon_i = x_{i}^T\beta_{i} + \varepsilon_i, \quad \beta_{i}=\lambda_{x_{i}}^{-1/2}\beta, \quad i=1, \dots,n,  $$
giving individual-specific regression coefficients, similar to a random effects model. Such random effects in a mixed model framework are typically assumed to follow a multivariate normal distribution and estimated using empirical Bayes methods, an approach which could also be suitable for fridge. 

Based on the plug-in approach to estimate the tuning parameter, a possible extension of fridge is to utilize external data to form the plug-in estimate. The integration of external data, so-call co-data, such as p-values from earlier studies, gene annotations or other prior knowledge \citep{tai2007incorporating,bergersen2011weighted}, is typically achieved by up- and down-weighting variables according to some measure of importance and the approach has also been studied in the setting of ridge regression \citep{wiel2016better}. Basing the plug-in estimates in the fridge framework additionally, or solely, on external data opens for new approach to utilizing co-data in integrative analyses. 

\bibliographystyle{chicago}
\bibliography{BibRidge}

\end{document}